\documentclass[12pt]{elsarticle}
\usepackage{graphicx}
\usepackage{epsfig}
\usepackage{subcaption}
\usepackage{amssymb,amsmath}
\usepackage{float}
\usepackage{algorithm2e}
\usepackage{soul,xcolor}
\usepackage{multirow}
\usepackage{array}
\usepackage{todo}
\usepackage{hyperref}
\usepackage{nicefrac} 
\usepackage{booktabs} 
\usepackage[percent]{overpic}
\usepackage[left=1in,right=1in,top=1in,bottom=1in]{geometry}

\renewcommand{\vec}[1]{\mathbf{#1}}

\usepackage{pgfplots} 
\usepackage{tikz}
\begin{document}
\begin{frontmatter}
  \title{Multiphase flows with compressible and incompressible phases}
  \author[mss]{Prapanch Nair\corref{cor1}}
  \ead{prapanch.nair@fau.de}
  \author[iisc]{Gaurav Tomar\corref{cor2}}
  \cortext[cor1]{Principal corresponding author}
  \address[mss]{Institute for Multiscale Simulation, Friedrich-Alexander Universit\"at Erlangen-N\"urnberg, Erlangen, Germany.}
  \address[iisc]{Department of Mechanical Engineering, Indian Insitute of Science, Bangalore, India.}

  \begin{abstract}
    Gas bubbles immersed in a liquid and flowing through a large pressure gradient undergoes volumetric deformation in addition to possible deviatoric deformation. While the high density liquid phase can be assumed to be an incompressible fluid, the gas phase needs to be modelled as a compressible fluid for such bubble flow problems. The Rayleigh--Plesset (RP) equation describes such a bubble undergoing volumetric deformation due to changes in pressure in the ambient incompressible fluid, assuming axisymmetric dynamics. We propose a compressible-incompressible coupling of Smoothed Particle Hydrodynamics (SPH) and validate this coupling against the RP model in two dimensions. For different density ratios, a sinusoidal pressure variation is applied to the ambient incompressible liquid and the response of the bubble is observed and compared with the solutions of the axisymmetric RP equation. 
  \end{abstract}
\end{frontmatter}

\section{Introduction}  
\label{intro}
The response of compressible gas bubbles flowing across a pressure gradient has 
a wide range of industrial applications. 
For example, effervescent flow atomizers (EFA) atomize liquids more efficiently 
than conventional atomizers by introducing gas bubbles into liquid jets. The gas bubbles 
expand and explode due to the pressure gradient across the nozzle and eventually 
cause fragmentation of
the surrounding liquid phase. 
Due to their insensitivity to liquid viscosity
\cite{loebker1997high} and reduced maintenance cost \cite{sovani2001effervescent},
EFAs have become a commonplace device to achieve efficient drop distribution and larger
spray cone angles \cite{whitlow1993effervescent}. 
So far, experimental studies have been helpful in only understanding 
the bulk properties of effervescent 
flows \cite{jedelsky2009development,Gadgil2011,avulapati2012experimental}. 
A spatially and temporally resolved  understanding of the compressible bubbles
deforming due to pressure changes  is generally lacking in experimental studies.

Numerical simulations where two phases with different compressibility 
treatments are required, on the 
other hand, face other difficulties. The following are the three most
important issues in numerical modeling of such two phase 
compressible-incompressible (CI) systems \cite{billaud2011simple}:
\begin{enumerate}
  \item Density is constant in the incompressible phase and is dependent on pressure and temperature in the compressible phase. 
  \item A zero divergence constraint has to be applied in the incompressible phase, and modeling this constraint near
    the interface involves numerical challenges. 
  \item The normal stress balance at the interface can become complicated due to pressure waves in the compressible
    phase.
\end{enumerate}
Hence numerical simulation approaches for solving CI 
two phase flow problems are rather limited. Traditional grid based CFD 
methods attempt such problems using either a 1) Unified approach or a 2) Non-uniform
approach. Unified approach uses the same solver in the entire domain with 
localization parameters that modify the
terms in the Governing Equations based on the phase of the node (
discretized element) in question. Such methods
traditionally assume weak compressibility (see \cite{hauke1994unified,hauke1998comparative}) 
in the incompressible phase and hence
compressible Navier--Stokes equation is solved in the entire domain. Such methods 
address the incompressibility as a limiting case. 
A major disadvantage of such an approach is the huge
constraint on the time steps ($\Delta t$) used in time integration, since 
the wave speeds used in the time step criteria
are high in the near-incompressible region. Several methods use similar approach together with
interface tracking schemes (for e.g., Low Mach Schemes, Generalized Projection Method, Stabilized Finite Element method,
Marker and Cell scheme generalized to Euler equations and the Discontinuous Galerkin scheme \cite{pesch2008discontinuous}). 
The `non-unified' schemes, 
on the other hand, solve both the fluid phases differently in a spatially decomposed 
domain. Explicit or implicit coupling is used to achieve stress balance. Such 
approaches are relatively less explored in literature \cite{hauke1998comparative}. 
One difficulty with such methods is the requirement to use
conservative formulation in the compressible phases, 
and non-conservative formulations based on primitive variables in the
incompressible phases. Separate set of governing equations 
require different numerical approaches making
the solution complex to implement.

We introduce a meshless method based on the Smoothed Particle Hydrodynamics (SPH) 
method, that couples a truly 
incompressible phase \cite{cummins1999sph} with a compressible phase. 
In 
most cases in literature (for example, \cite{ferrari2011high,liu2005ghost}), a 
stiff equation of state is used to obtain pressure in the nearly incompressible 
phase. Numerically, 
such approaches result in severe time-step 
restrictions \cite{violeau2014maximum}. 
A more strict 
incompressibility condition could be  achieved by solving the pressure Poisson 
equation (PPE) \cite{cummins1999sph} or a condition for isochoricity \cite{nair2015volume}. 
Mass conservation across interfaces in naturally satisfied due to the use of
constant mass particles in SPH. Hence, the coupling of phases with different 
compressibility treatments is feasible \cite{lind2016incompressible}. This paper describes the 
coupling of compressible SPH with truly incompressible SPH, focusing on applications
to effervescent flows where a compressible gas is immersed in an incompressible 
fluid. 

This paper is organized as follows. In the following section 
we present the SPH implementation of the CI two phase flow simulation (CI--SPH).
In Sec. \ref{sec:rp} 
we derive the Rayleigh--Plesset (RP) problem applied to two dimensional finite 
domains of circular and rectangular shapes
and proceed to validate the CI--SPH method against the theory in Secs. \ref{sec:rp_rect}
and \ref{sec:rp_circ}. We compare the time response of a compressible bubble against
solutions of the RP equation and present the motion of the interface. 

\section{SPH for Compressible--Incompressible flows}
\label{sec:sph}
In what follows, we describe the SPH method used to couple the compressible
and incompressible phases and the method of application of time varying ambient 
pressure, $P_\infty(t)$.
The momentum conservation equation for the entire domain is given by
\begin{equation}
  \begin{split}                                                                 
    \left. \frac{d \vec{u} }{d t}\right| _a = & - \sum_b m_b\frac{p_a + p_b}{\rho_a \rho_b}  \nabla_a W_{ab}  \\
                                  &  + \sum_b m_b  \frac{\mu_a + \mu_b }{\rho_a \rho_b}F_{ab} \vec{u}_{ab} + \vec{f}^{\textrm{int}}_{ab} +\vec{f}_a^B,
  \end{split} 
      \label{eq:sphdiscretization}
\end{equation}
where $m$ is the mass, $\rho$ is the density, $p$ is the hydrodynamic pressure, $\mu$ is the coefficient of viscosity and $\vec{u}$ is the velocity at an SPH particle 
$a$ or SPH particle $b$ in the neighborhood of $a$. 
The function $W$ is the SPH smoothing function
(also known as the smoothing kernel) with a finite cut-off radius 
 defined for an SPH particle pair as $W_{ab} = W(r_{ab},h)$, where $h$ is 
the smoothing length of the kernel. The gradient of the smoothing function 
appears as $\nabla_a W_{ab}$ for an SPH particle $a$ with respect to its 
neighbor $b$. The radial derivative of the kernel, given by 
$F_{ab}$ \cite{monaghan1992smoothed} is computed from the gradient of $W$ as 
\begin{equation}
  F_{ab}= \frac{\vec{r}_{ab} \cdot \nabla_a W_{ab}}{r_{ab}^{2}+\epsilon^2} 
\end{equation}
where $\epsilon$ is a small number introduced to avoid division by zero in the rare
event of SPH particles overlapping in position and is usually set to $(0.01h)^2$.

The incompressible phase is bounded by the interface with the compressible
phase on one side and the free surface on the other (see Figs. \ref{fig:rp-1d} and \ref{fig:rp}).
The PPE that ensures incompressibility and accounts for a Dirichlet BC applied through the 
truncated domain at the free surface \cite{nair2014improved} is given by 
 \begin{equation}                                                 
   (p_a - P_\infty)  \kappa - \sum_{b} \frac{m_b}{\rho_b}\frac{4p_b}{\rho_a +\rho_b} F_{ab}
     = \sum_{b}\frac{m_b}{\rho_b}\left(    \frac{\vec{u}_{ab} \cdot \nabla_a W_{ab}}{\Delta t} - \frac{4P_\infty}{\rho_a +\rho_b}F_{ab} \right).
               \label{eq:freesurf}                                                            
 \end{equation}
This equation is solved using a linear solver such as BiCGSTAB \cite{sleijpen1994bicgstab}, for 
the unknown pressures $p_a$.
   Here $P_\infty$ represents the time varying ambient pressure that is applied
   at the free surface. The term  $\kappa $ in the above equation, given by    
   \begin{equation}                                                             
     \kappa = \sum_{b_{\textrm{bulk}}}\frac{m_b}{\rho_b} \frac{4}{\rho_a +\rho_b}\frac{\vec{r}_{ab} \cdot        
     \nabla_a W_{ab}}{r_{ab}^{2}+\epsilon^2}  ,
     \label{eq:kappa}
   \end{equation}
is a factor which remains constant for a given domain with given smoothing 
parameters and constant density. 
The particles in the compressible region form the Dirichlet boundary condition for the solver
on the inner boundary of the incompressible domain. 

For the compressible phase, we consider a linear isothermal pressure-density relation given by 
\begin{equation}
  p_i =  \rho_i c^2.
\end{equation}
The velocities encountered are  subsonic in all the cases considered here. 
The speed of sound in the compressible phase is denoted by $c$. In order to 
correspond to the RP equations we switch off the viscous forces in the compressible
phase. For successful application of the time varying pressure at the ambience,
the pressure gradient term also needs to account for deficiency near the free surface. 
Since pressure at the free surface (Dirichlet BC for pressure) varies in time,
the pressure gradient computation should account for it as a time varying
boundary condition. We implement a penalty approach similar to that used in 
the solution for PPE to set this Dirichlet BC. Assuming, $\rho_a = \rho_b$ since
the free surface only occurs in the incompressible phase, we write
\begin{equation}
  \vec{F}^p_a = - \sum_b m_b\frac{p_a + p_b}{\rho_a \rho_b}  \nabla_a W_{ab} - \frac{(P_\infty + p_a)}{\rho_a} \left(0.0 - \sum_b \frac{m_b}{\rho_b} \vec{\nabla W_{ab}} \right),
\end{equation}
where $\vec{F}^p_a$ is the pressure gradient term that is to replace the first 
term on the right hand side of Eq. \ref{eq:sphdiscretization}, in the incompressible
phase near the free surface. Thus the time 
varying ambient pressure BC can be
applied to both the pressure solver
as well as the pressure gradient 
operator for the incompressible phase.

Surface tension force $\vec{f}_{ab}^\mathrm{int}$ can be applied at the 
two phase interface using any of the surface tension models available for 
SPH in literature (for example, \cite{rezavand2018isph} and \cite{adami2010new}). We use a simple model based on 
the continuum surface force (CSF) that is based on computation of gradients
of a discontinuous color function across the two phase interface \cite{morris2000simulating}.
For details on this CSF implementation please see \cite{morris2000simulating}.

\section{Validation of the CI-SPH}
\label{sec:validation}
\begin{figure}[htb]
  \begin{center}
    \begin{subfigure}[b]{2in}
      \includegraphics[width=\textwidth]{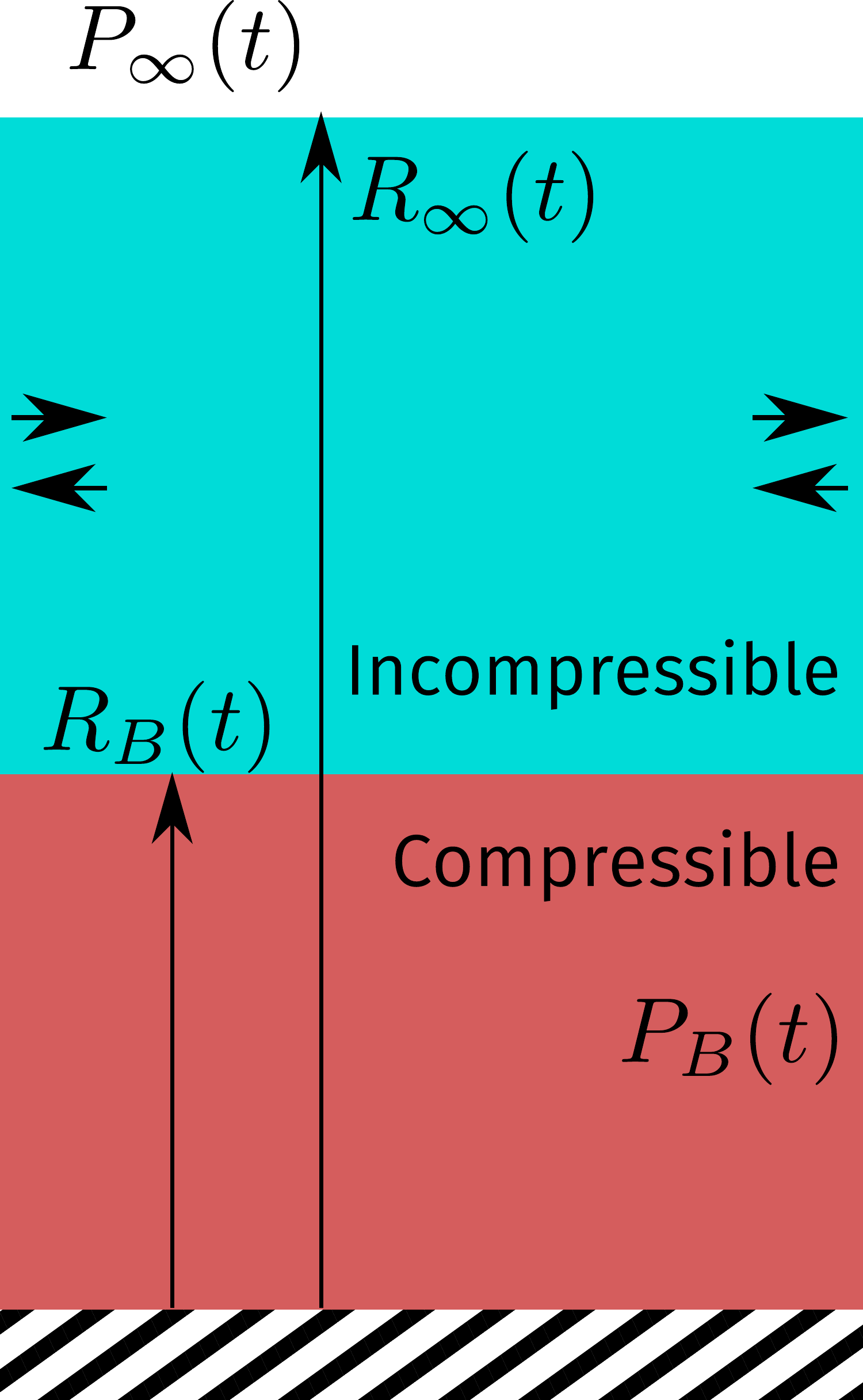}
      \caption{ 1D.}
      \label{fig:rp-1d}
    \end{subfigure}
    \hspace{0.25in}
    \begin{subfigure}[b]{3in}
      \includegraphics[width=\textwidth]{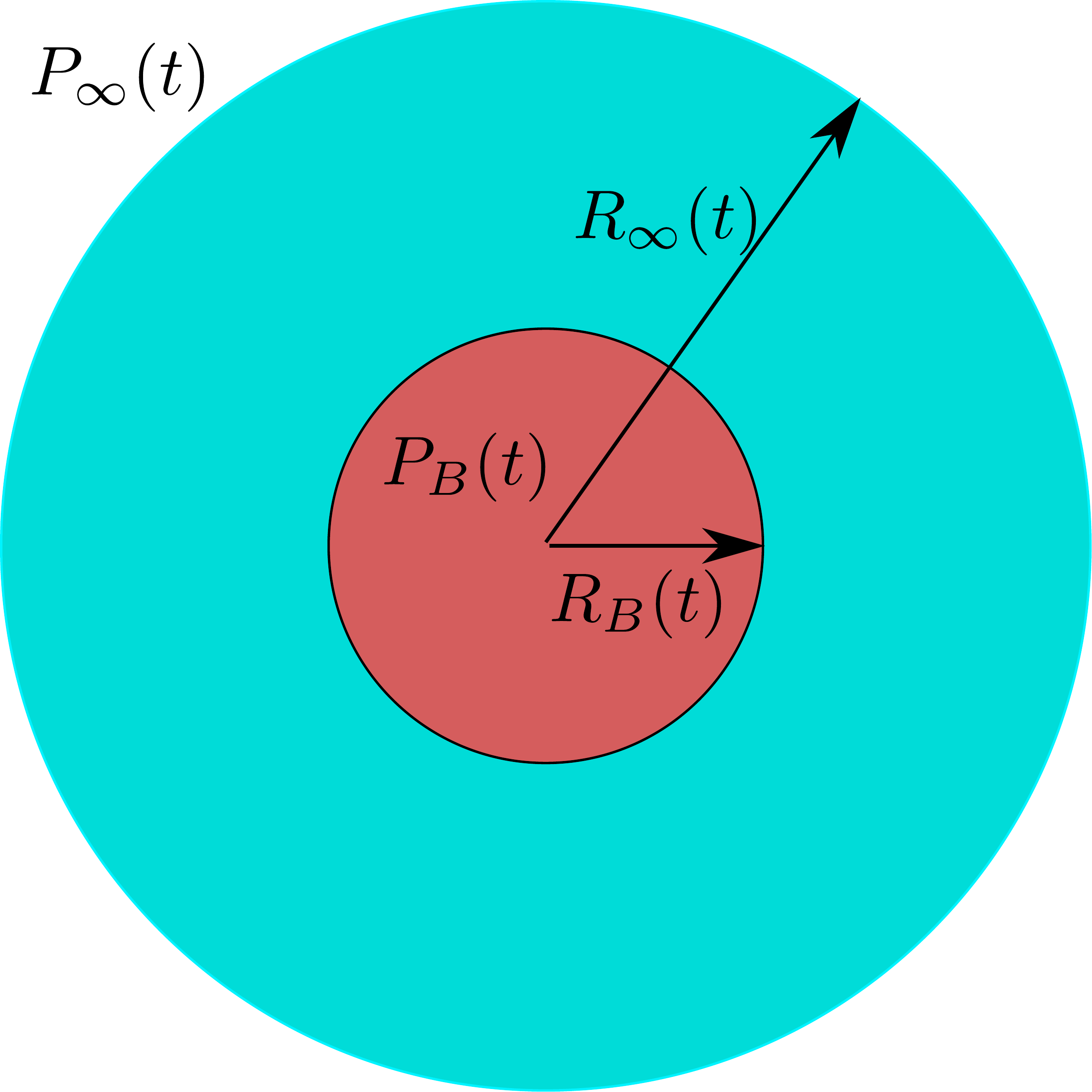}
      \caption{2D}
      \label{fig:rp}
    \end{subfigure}
    \caption{Schematic of the Rayleigh-Plesset problem in 2D}
    \label{schematic}
  \end{center}
\end{figure}
The RP equation in 2D is derived in this section for a rectangular and a circular 
geometry. We then preform the validation of the above introduced CI-SPH method to solve the 
RP equation in rectangular and circular  domains as depicted in Fig. \ref{schematic}.

\subsection{Rayleigh-Plesset equation in two dimensions}
\label{sec:rp}
The RP equation governs the radial deformation of a (compressible) 
spherical bubble in an infinite body of incompressible liquid \cite{lin2002inertially,plesset1949dynamics,rayleigh1917viii}. In its general 3D form (for a spherical bubble) the 
equation is given by 
\begin{equation}
  \frac{P_B(t)-P_\infty(t)}{\rho_L} = R\ddot{R} + \frac{3}{2}\left(\dot{R}\right)^2 + \frac{4\nu_L}{R}\dot{R} + \frac{2S}{\rho_LR},
    \label{eq:rp}
\end{equation}
    where
    $ P_B(t)$ is the uniform pressure within in the bubble (and is function of 
    density and internal energy, in general), 
    $P_\infty(t)$ is the external pressure far from the bubble that varies with time $t$,
    $\rho_L$ is the density of the surrounding liquid,
    $R(t)$ is the radius of the bubble,
    $\nu_L$ is the kinematic viscosity of the liquid and
    $S$ is the surface tension of the bubble.

    For a given $P_\infty(t)$, and known initial bubble pressure, $P_B(t=0)$, 
    the RP equation can be used to solve for the bubble radius 
    $R(t)$. 
    For a finite annular spherical domain of the incompressible 
    fluid with an outer radius $R_\infty$, the RP equation can be written as 
    \begin{equation}
      \left[ \left( \frac{P_B -P_\infty}{\rho_L} - \frac{2S}{\rho_L R}-
                           \left(\frac{R^4}{R_\infty ^4} -1
                   \right)\frac{\dot{R}^2}{2}\right)\frac{R_\infty}{R_\infty - R}  
                   - 2 \dot{R}^2 \right]\frac{1}{R}  = \ddot{R},
      \label{eq:rp3d}
    \end{equation}
    where $R_\infty$ represents the outer radius of the drop of incompressible fluid which in turn
    holds the compressible bubble at its center. 

    The 2D version of RP equation can be derived as follows.
    From conservation of momentum we have,
    \begin{equation}
      \frac{\partial u}{\partial t} + u \frac{\partial u}{\partial r} = - \frac{1}{\rho}\frac{\partial p}{\partial r} \\
    \end{equation}
where $r$ is the radial coordinate, $u$ is the velocity of the interface and $R$ is the radial position of the interface of the compressible bubble. From the requirement for continuity, the radial velocity at the interface $dR/dt$ and the velocity at any point in the fluid can be related as
\begin{equation}
  u = \frac{R}{r}\frac{dR}{dt} .
\end{equation}
Therefore,
\begin{align}
  -\frac{1}{\rho}\frac{\partial p}{\partial r} &= \frac{\partial}{\partial t}\left(
  \frac{R}{r}\frac{dR}{dt}\right) + \frac{R}{r}\frac{dR}{dt}\frac{\partial}{\partial r} \left( \frac{R}{r}\frac{dR}{dt}\right) \\
  & = \frac{1}{r}\left(R\frac{d^2 R}{dt^2} + \left( \frac{dR}{dt}\right)^2\right) - \frac{R^2}{r^3}\left( \frac{dR}{dt}\right)^2 
\end{align}
Integrating this equation by applying the limits from a point just outside the interface (excludes surface tension effects) to the outer boundary of the incompressible fluid, 
\begin{align}
  - \int \limits_{p_B'} ^{p_\infty}  \frac{1}{\rho}\frac{\partial p}{\partial r} &= \int \limits_R^{R_{\infty}}\frac{1}{r}\left(R\frac{d^2 R}{dt^2} + \left( \frac{dR}{dt}\right)^2\right) - \frac{R^2}{r^3}\left( \frac{dR}{dt}\right)^2\\ 
    \implies \ddot{R} &= \left( \frac{p_B-p_{\infty}}{\rho} + \dot{R}^2\left( \frac{1}{2}\left(1-\left(\frac{R}{R_{\infty}}\right)^2\right) - \ln \frac{R_{\infty}}{R} \right) \right) \frac{1}{R\ln \left( \frac{R_{\infty}}{R}\right)}
\end{align}
Here the pressure $p_B$ is accounted for right outside the interface. If $P_B$ is the pressure
within the bubble, then considering the Young--Laplace pressure jump due to 
surface tension $P_B-p_B=S /R$, where $S$ is the surface tension coefficient, the above expression can be written as,
\begin{equation}
  \left[  \frac{P_B -P_\infty}{\rho_L} - \frac{S}{\rho_L R}- 
    \left(\frac{R^2}{R_\infty ^2}  -1
    \right)\frac{\dot{R}^2}{2}  - 
    \dot{R}^2\ln \frac{R_\infty}{R}
  \right]\frac{1}{R\ln \frac{R_\infty}{R}}  = \ddot{R},
  \label{eq:rp-2d}
\end{equation}
which is the 2D equivalent of the RP equation for a `circular' bubble (see Fig. \ref{schematic}b).
For a rectangular bubble (see Fig. \ref{schematic}a) this equation may be derived similarly and is given by 
\begin{equation}
  \left( \frac{P_B -P_\infty}{\rho_L}\right)\frac{1}{R_\infty-R} = \ddot{R}
  \label{eq:rp-1d}
\end{equation}
In this paper the circular and rectangular versions of RP equations (Eqs. \ref{eq:rp-2d} and \ref{eq:rp-1d}) 
are solved numerically to obtain the time response of the compressible bubble. 
Throughout this paper we use the explicit $4$th order Runge--Kutta method 
to solve the RP equations.    
\begin{figure}[h!]
  \begin{center}
    \begin{subfigure}[b]{4.5in}
      \includegraphics[width=\textwidth]{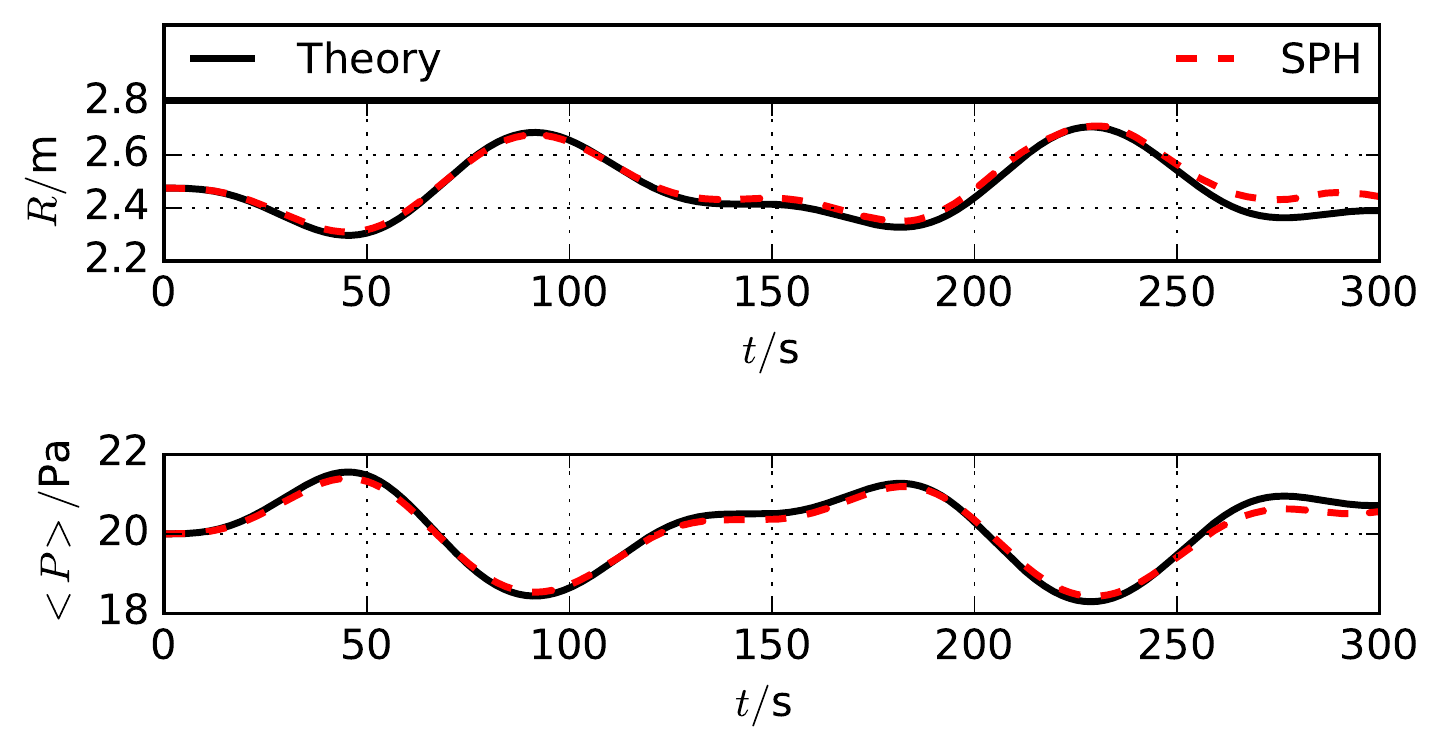}
      \caption{Density Ratio 10:1}
      \label{dr10_1d}
    \end{subfigure}\\
    \begin{subfigure}[b]{4.5in}
      \includegraphics[width=\textwidth]{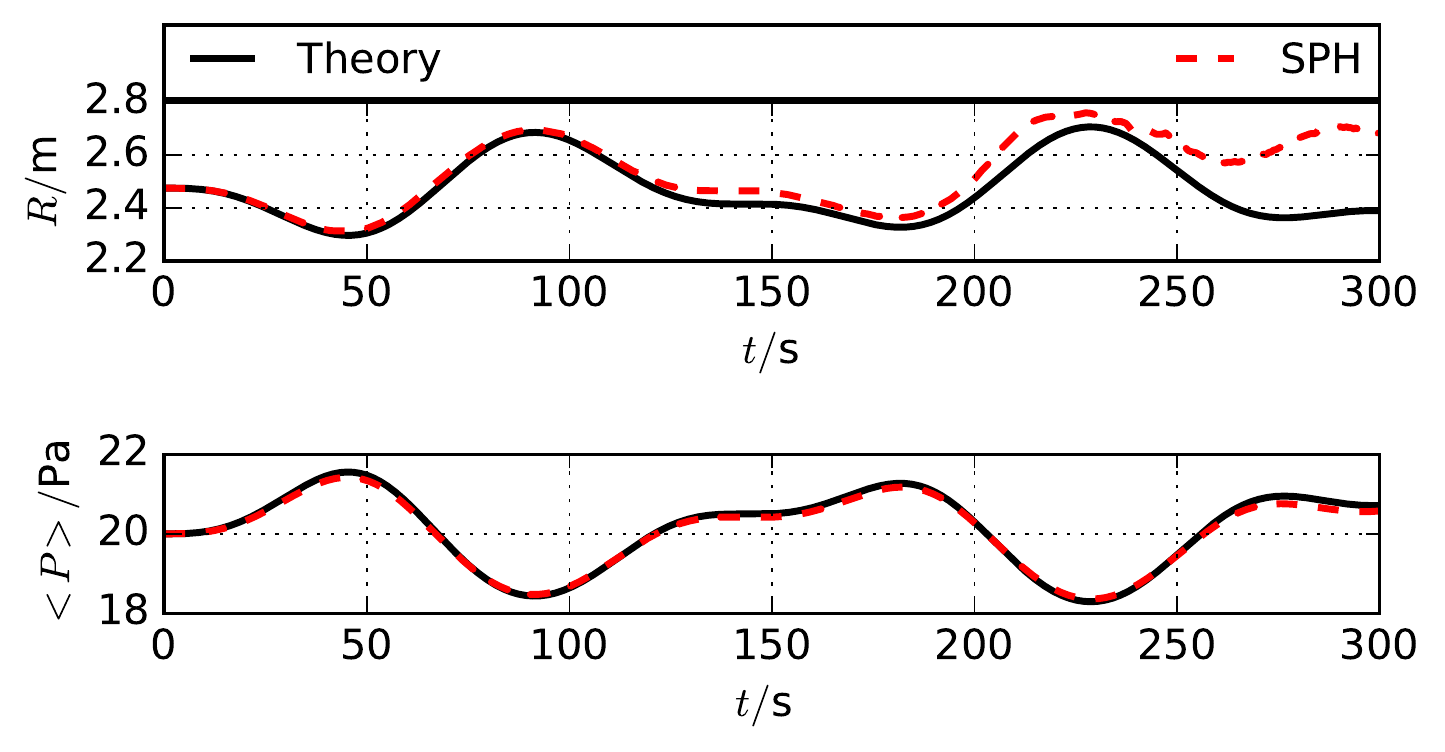}
      \caption{Density Ratio 100:1}
      \label{dr100_1d}
    \end{subfigure}\\
    \begin{subfigure}[b]{4.5in}
      \includegraphics[width=\textwidth]{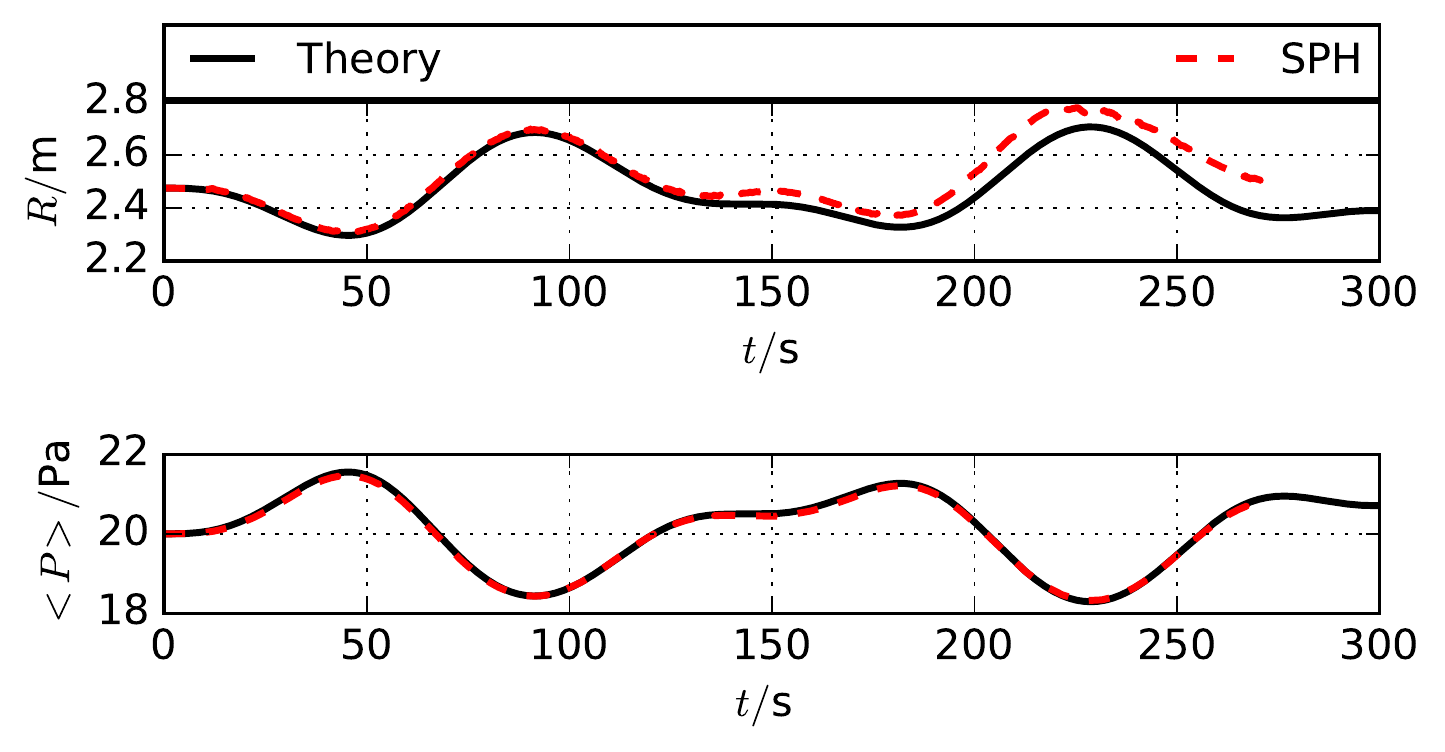}
      \caption{Density Ratio 1000:1}
      \label{dr1000_1d}
    \end{subfigure}
    \caption{Pressure and radius variation with time for different density ratios. Comparison with numerical solution of Rayleigh--Plesset equation.}
    \label{response_rect}
  \end{center}
\end{figure}

\begin{figure}[htb]
  \begin{center}
    \begin{subfigure}[b]{0.32\textwidth}
      \includegraphics[width=\textwidth]{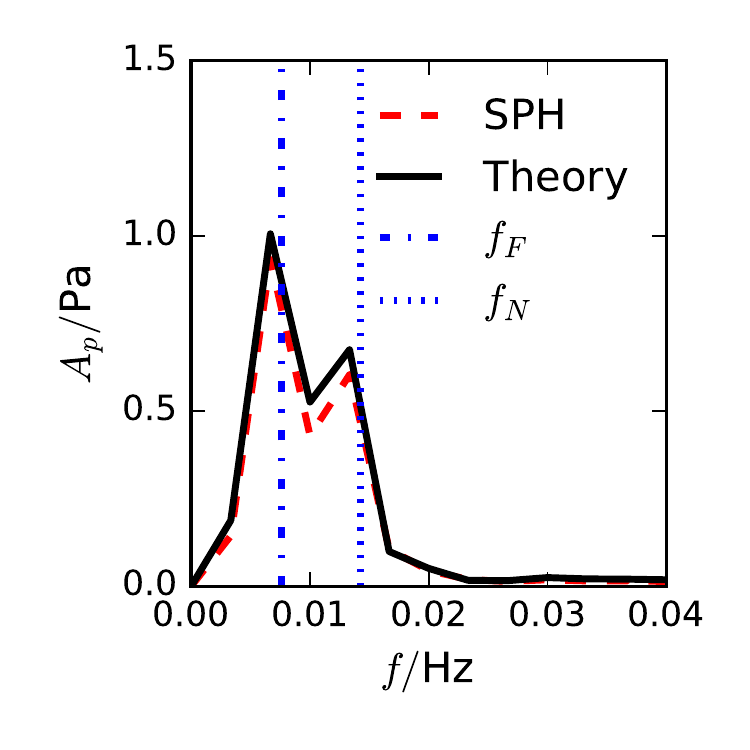}
      \caption{Density Ratio 10:1}
      \label{dr10_1df}
    \end{subfigure}
    \begin{subfigure}[b]{0.32\textwidth}
      \includegraphics[width=\textwidth]{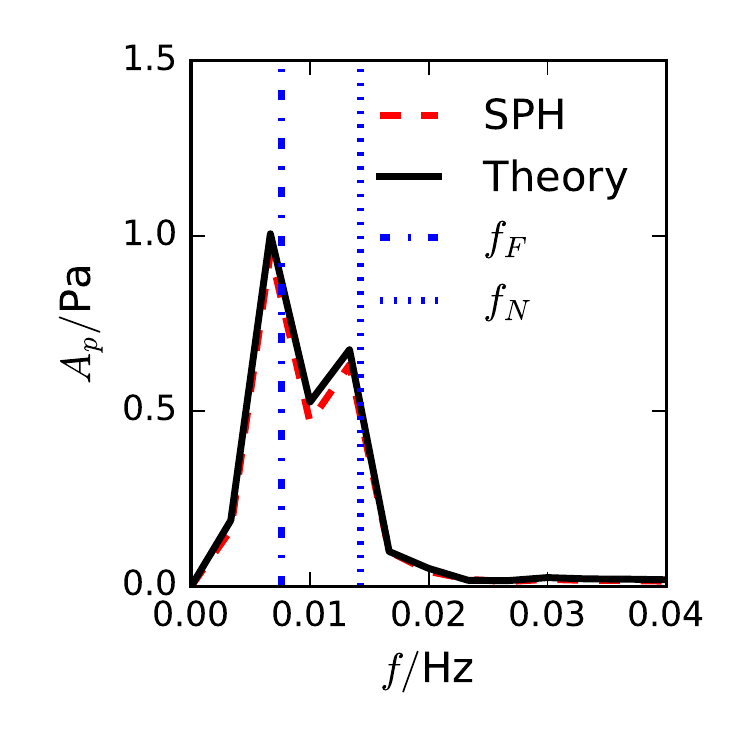}
      \caption{Density Ratio 100:1}
      \label{dr100_1df}
    \end{subfigure}
    \begin{subfigure}[b]{0.32\textwidth}
      \includegraphics[width=\textwidth]{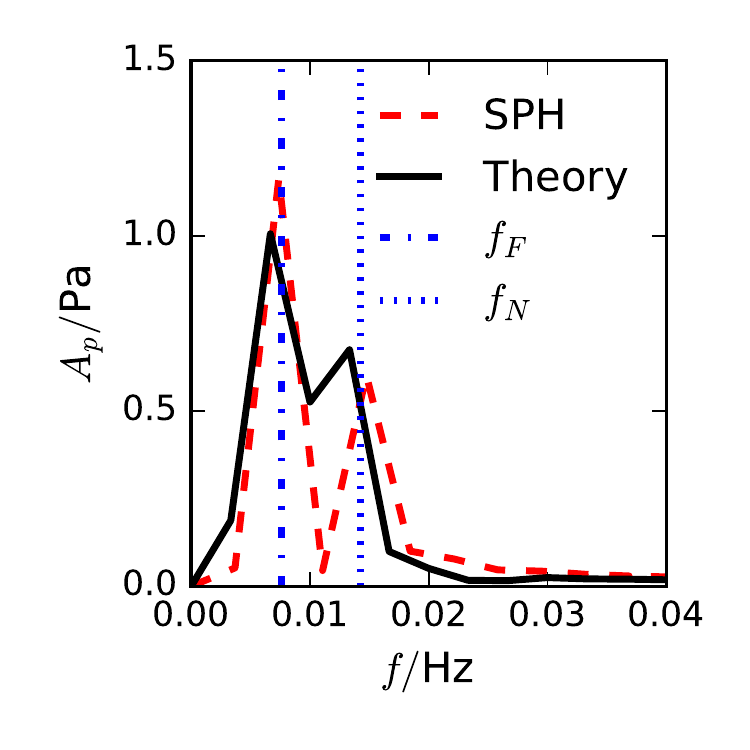}
      \caption{Density Ratio 1000:1}
      \label{dr1000_1df}
    \end{subfigure}
    \caption{Discrete Fourier transform of the response of the pressure of the bubble for different density ratios. Comparison with the numerical solution of Rayleigh--Plesset equation. The analytical values for the forcing frequency $f_F$ (dotted vertical line) and natural frequency $f_N$ (dash-dotted vertical line) are labelled in (c).}
    \label{frequency_rect}
  \end{center}
\end{figure}
\begin{figure}[htb]
  \begin{center}
    \begin{subfigure}[b]{0.2\textwidth}
      \includegraphics[width=\textwidth,trim={0in 0in 0in 8in},clip]{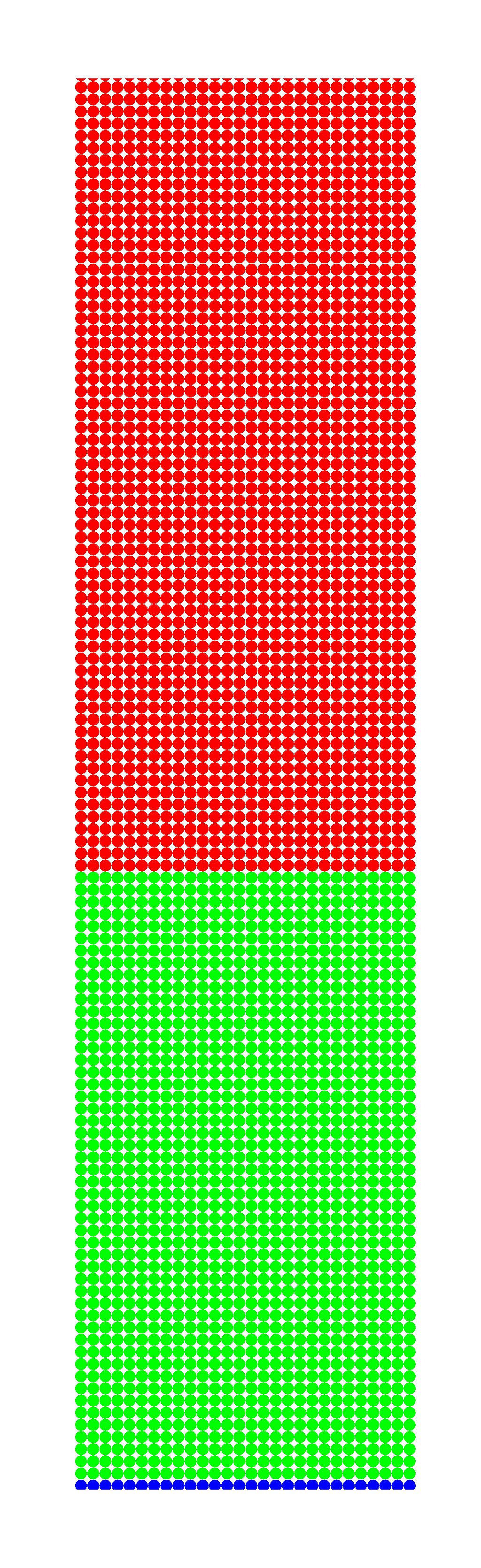}
      \caption{t=0}
      \label{dr10_0}
    \end{subfigure}
    \begin{subfigure}[b]{0.2\textwidth}
      \includegraphics[width=\textwidth,trim={0in 0in 0in 8in},clip]{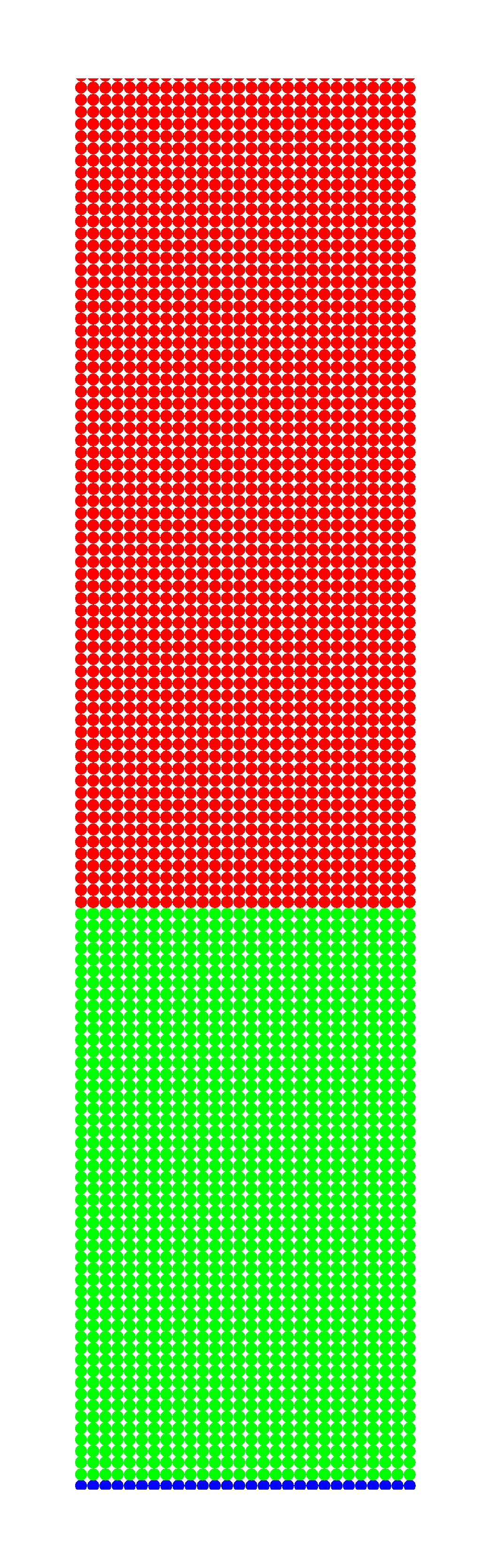}
      \caption{t=50}
      \label{dr10_50}
    \end{subfigure}
    \begin{subfigure}[b]{0.2\textwidth}
      \includegraphics[width=\textwidth,trim={0in 0in 0in 8in},clip]{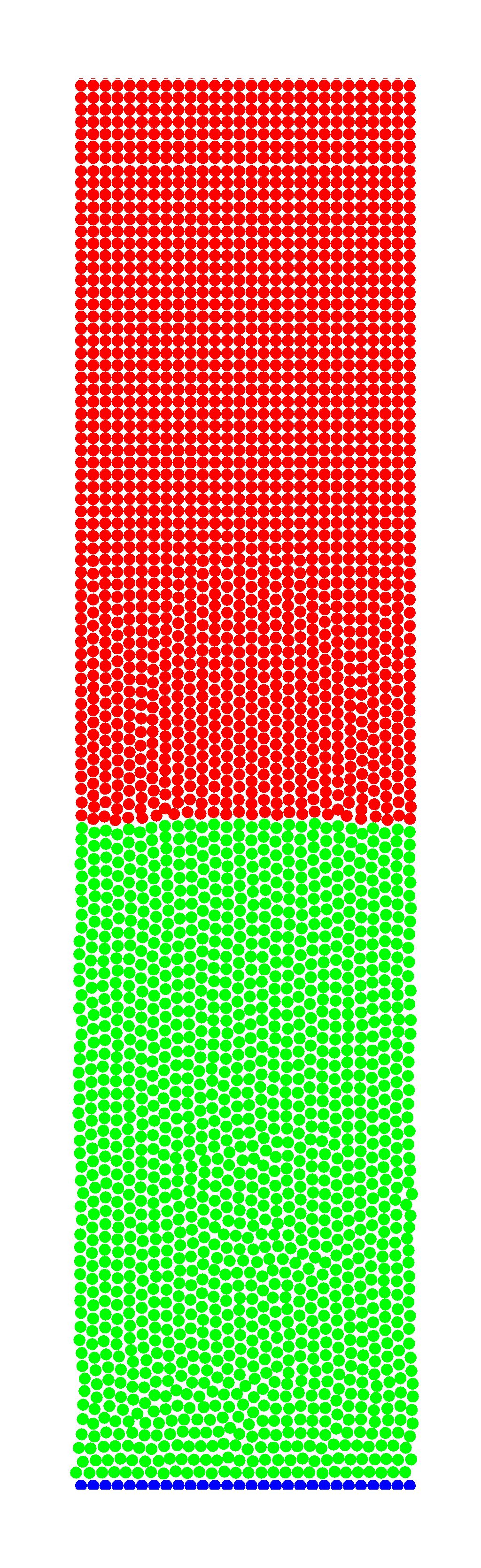}
      \caption{t=225}
      \label{dr10_225}
    \end{subfigure}
    \begin{subfigure}[b]{0.2\textwidth}
      \includegraphics[width=\textwidth,trim={0in 0in 0in 8in},clip]{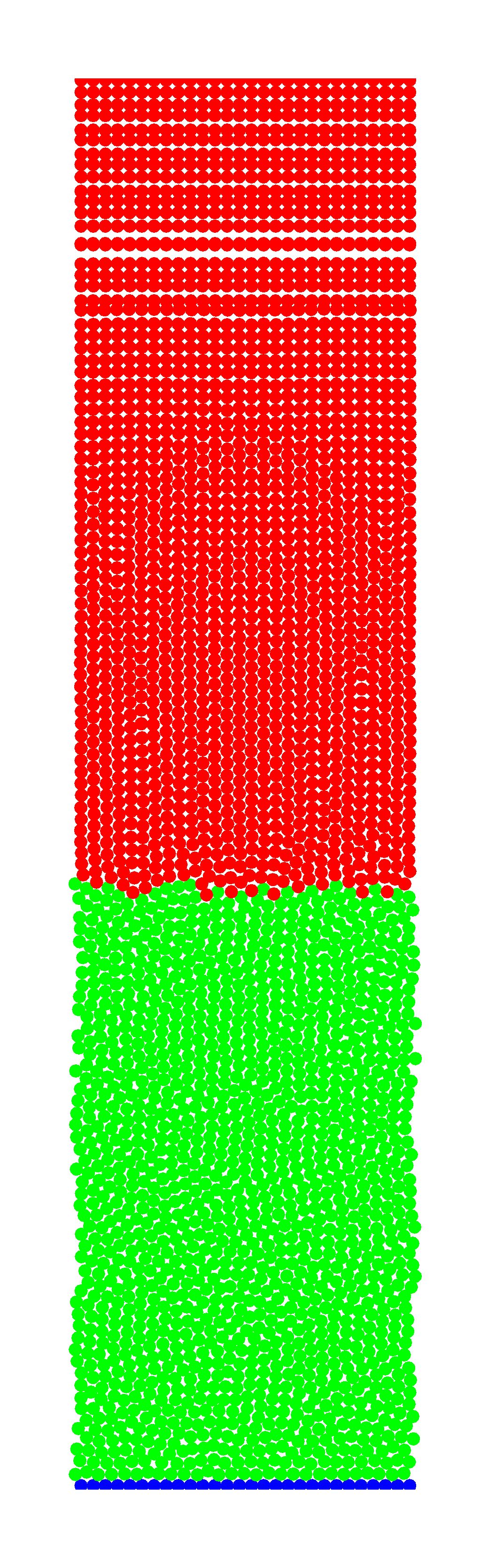}
      \caption{t=285}
      \label{dr10_285}
    \end{subfigure}
    \caption{Particle configuration of the compressible (green) and incompressible (red) phases, corresponding to a density ration of 10:1.}
    \label{rect_images}
  \end{center}
\end{figure}
\subsection{Rectangular domain}
\label{sec:rp_rect}
Simulations of the RP problem for a rectangular bubble are presented here. The compressible phase 
is bounded by a wall at the bottom and the incompressible phase at the top. Though the dynamics 
are confined to one dimension, we use a 2 dimensional domain with a finite width and periodic boundary conditions at the vertical edges of the domain.  
In the rectangular case we have neglected the surface tension. 
The rectangular bubble 
has an initial radius (distance of interface from the bottom wall) of $2.5$ units and $R_\infty$ is $6$ units. The following three different density ratios have been considered 1000:1, 100:1 and 10:1 and the initial pressure inside the bubble is set to be at $P_B(t=0)=\rho c^2 = 20$ in each case. The value of speed of sound was changed correspondingly for each density ratio. A time varying pressure given by
\begin{equation}
  P_\infty (t) = 20+1.5\sin \left( \frac{2\pi t}{70}\right)
\end{equation}
 is applied at the free surface of the incompressible fluid. Initially the 
 pressure within the bubble is the same as that in the outer boundary. As the pressure increases, 
 the volume of the bubble decreases and its pressure increases, while the C--I interface
 moves towards the bottom wall. 

 The natural frequency for the 1D bubble can be derived from a linear stability analysis. 
 Assuming the initial pressure in the bubble $p_{B0}$ to be the same as the 
 far-field pressure outside of the incompressible region, we have
 \begin{equation}
   \frac{P_B - P_{\infty}}{\rho_L H} = \ddot{R}.
 \end{equation}
 Here $\rho_L$ is the density of the incompressible fluid and $H = R_{\infty} - R$. 
 Subscript is added to distinguish it from the density in the bubble, $\rho_B$.
 Let $P_{B0} = \rho_{B0}c^2$ at $t=0$. Let $P_B $ be the pressure at time $t$ 
 due to a small perturbation in $R$. For a small time after $t=0$,  
 $R = R_0 + r'e^{i\omega t}$, where $r'$ is the small perturbation given to $R$.
 \begin{align}
   \ddot{R} &= \frac{\rho_B c^2 - \rho_{B0}c^2}{\rho_L H} \\
       & = \frac{\rho_{B0}c^2}{\rho_L H} \left( \frac{R_0}{R} -1 \right) \nonumber \\
       & = \frac{\rho_{B0}c^2}{\rho_L H}\left(  \frac{R_0}{R_0 + r'e^{i\omega t}} -1  \right) \nonumber\\
       & = \frac{\rho_{B0}c^2}{\rho_L H}\left(   -\frac{r'}{R_0}e^{i\omega t}  \right) \qquad \text{(using binomial theorem )} \nonumber\\
   \implies \omega^2 &= \frac{\rho_B c^2}{H\rho_L R_0}
 \end{align}
 Thus the natural frequency of the rectangular RP system is given by
 \begin{equation}
   f_N = \frac{\omega}{2\pi} = \frac{1}{2\pi}\sqrt{\frac{\rho_{B0}c^2}{\rho_L H R_0}} \qquad,
   \label{1d_freqency}
 \end{equation}
 where $\rho_{B0}$ is the initial density in the bubble, $\rho_L$ is the density of the incompressible fluid, $H$ the distance between the CI interface and the free interface, $R_0$ is the initial radius of the bubble. 

The pressure and radius of the compressible 1D bubble is shown in Fig. \ref{response_rect}, for different density ratios. The simulation proceeds well during the first cycle of oscillation in all the cases.
 As the density ratio increases, the simulation approaches the limit of a 
 massless compressible bubble governed by the RP equation, as seen
 in the pressure plots in Fig. \ref{response_rect}. 
However, towards the end of  the first 
 cycle of 
 oscillations, the interface fails to remain stable, due to the absence of interface
 forces. This waviness in the interface reflects in the radius response as a deviation from the 
 theoretical response curve. On the other hand the pressure is computed as the mean of 
 the pressure of all particles within the bubble. 
 Hence, with increasing density ratio the response of pressure is captured 
 with increasing accuracy. 
 Figure \ref{frequency_rect} shows the frequency domain of the response of pressure 
 within the bubble. The natural frequency of the compressible bubble, as computed
 in Eq. \ref{1d_freqency} and the forced frequency of the applied pressure are 
 marked by blue lines. For all the three density ratios shown, we see that the 
 natural frequency and force frequency show up in the pressure response. The forcing
 frequency corresponds to the larger amplitude. The RP  theory,
 SPH simulations and the estimated frequencies compare well. 

Figure \ref{rect_images} shows the particle configuration of the compressible and incompressible phases for a density ratio of 10:1
 at different times in the simulation. The phase in the upper part is the 
 incompressible phase. At time $t=225$ the interface has become unstable and 
 is seen as a wavy line. However, the particle arrangement at earlier times is stable and is also reflected in the pressure and radius response in Fig. \ref{response_rect}. We would like to note here that these simulation were performed in the absence of any stabilizing effects due to surface tension forces at the interface.

 \subsection{Circular domain}
\label{sec:rp_circ}
 \begin{figure}[htb]
   \begin{center}
     \includegraphics[width=4.5in]{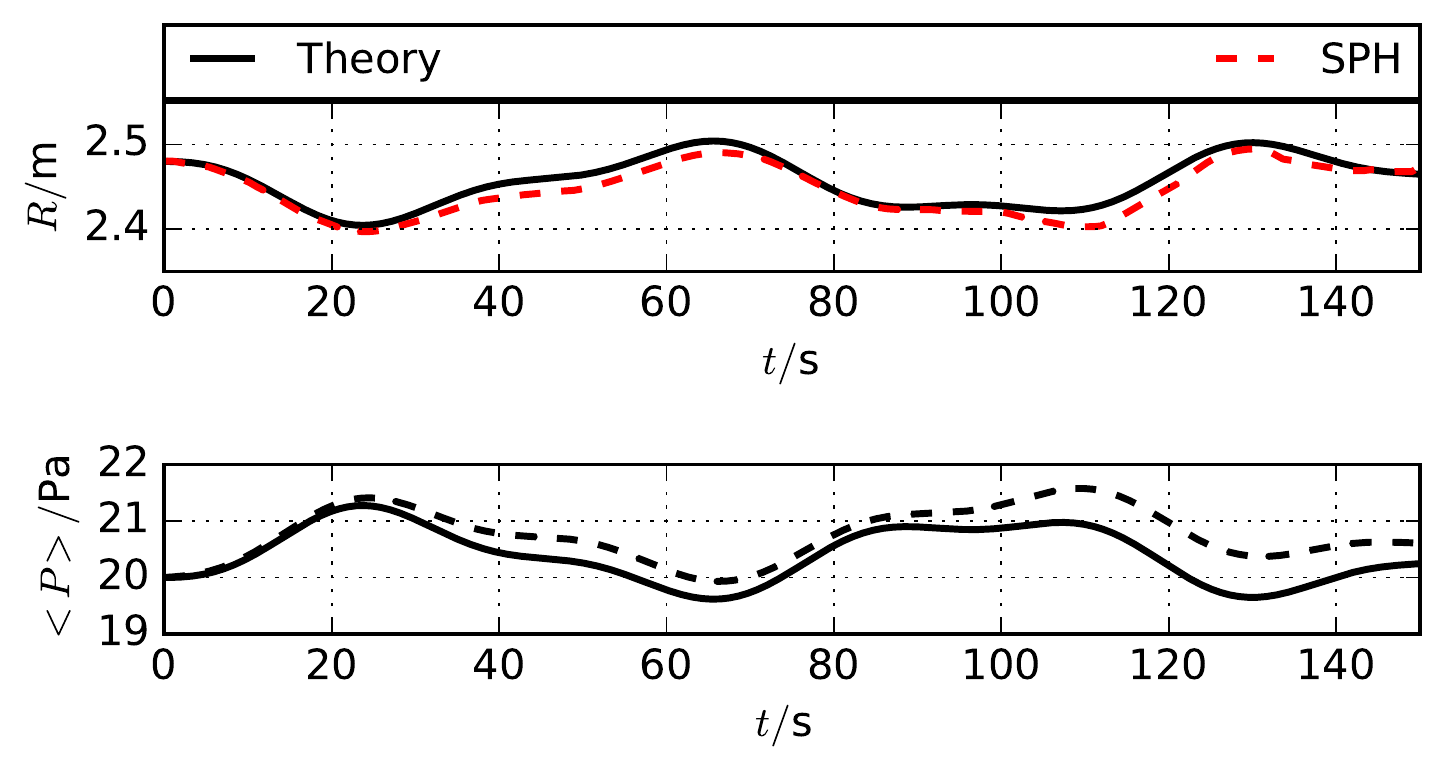}
     \caption{Radius and pressure response of the 2D drop in incompressible drop.}
     \label{fig:2drp}
   \end{center}
 \end{figure}
 \begin{figure}[htb]
   \begin{center}
     \includegraphics[width=3.5in]{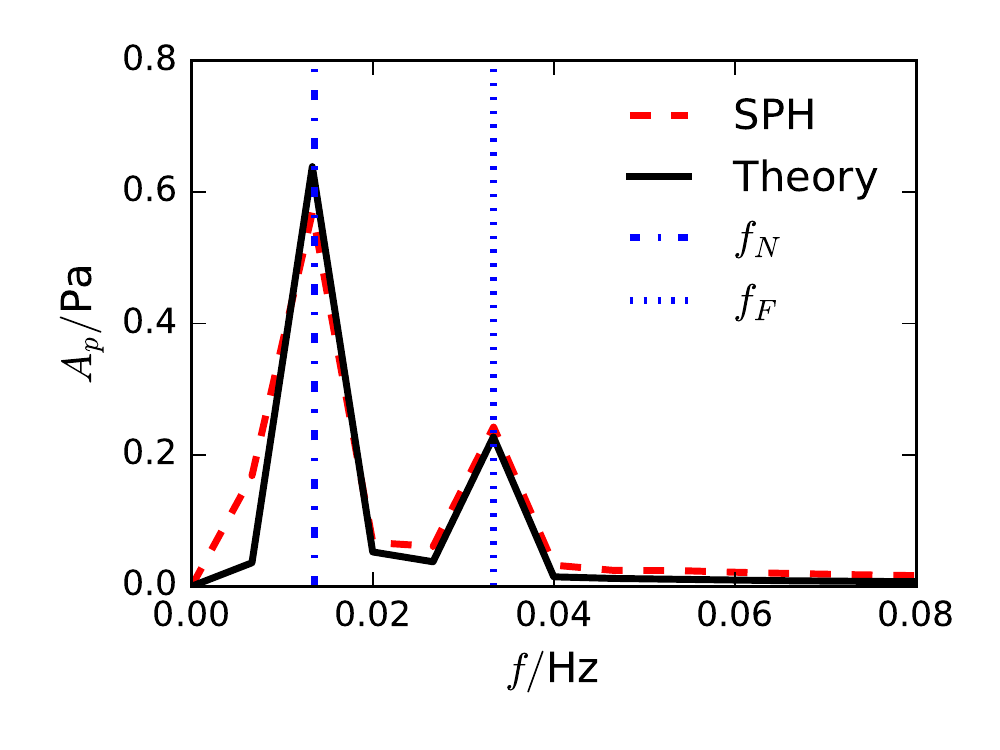}
     \caption{Frequency domain of the pressure of the 2D drop in incompressible drop.}
     \label{fig:2drpf}
   \end{center}
 \end{figure}
 \begin{figure}[htb]
   \begin{center}
     \includegraphics[width=5in]{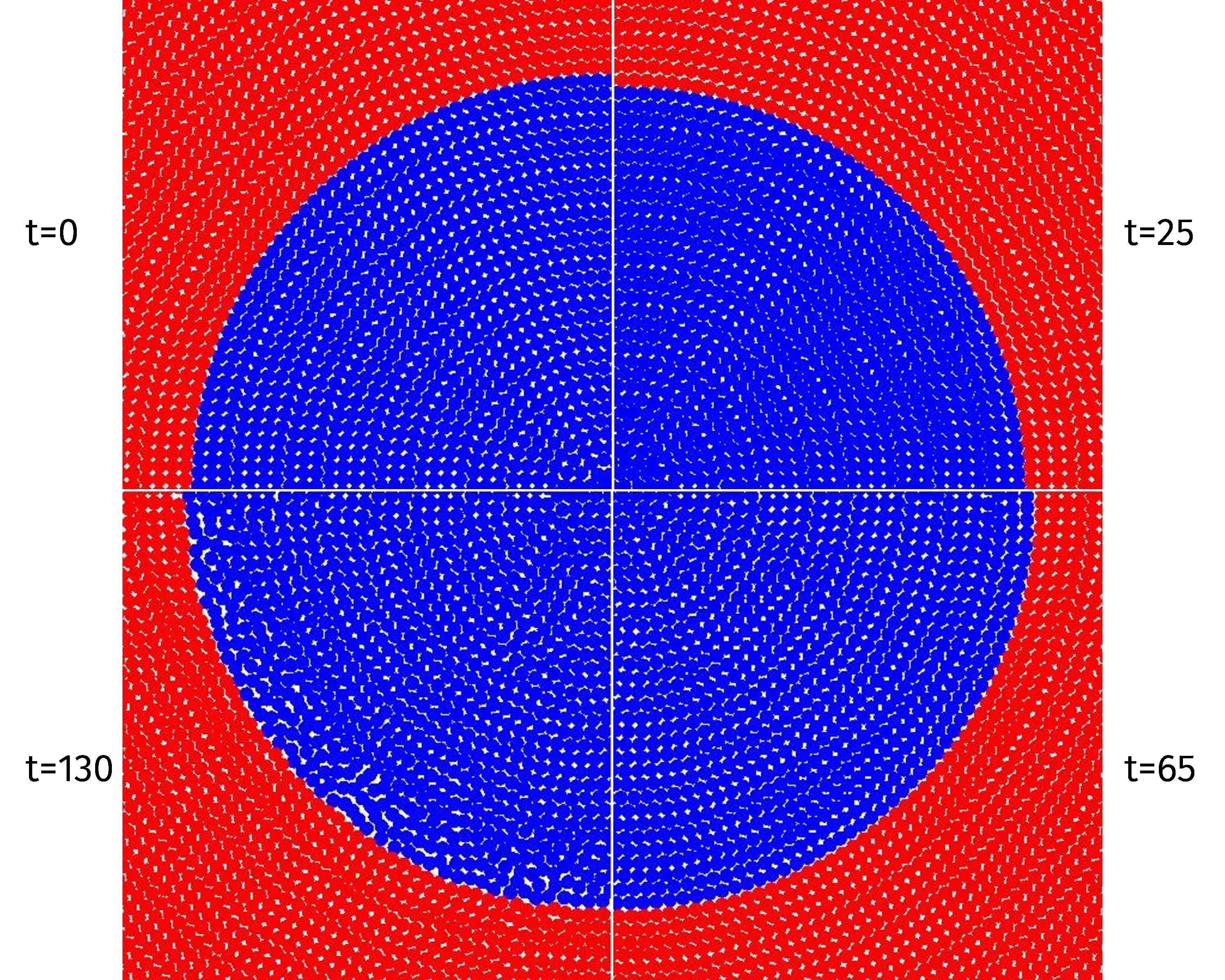}
     \caption{Position of interface at different times}
     \label{circle_time}
   \end{center}
 \end{figure}

 For the circular bubble case considered here (see Fig. \ref{schematic}), we  solve the RP 
 equation with an applied surface tension coefficient of $S=1$. The surface 
 tension model is based on the CSF model and is implemented as described in \cite{morris2000simulating}. 
 The time 
 varying pressure applied at the far-field boundary of the incompressible fluid is 
 \begin{equation}
   P_\infty (t) = 20+1.0\sin \left( \frac{2\pi t}{30}\right)
   \label{pressure_2d_applied}
 \end{equation}
 The corresponding natural frequency can be derived using a similar procedure 
 as with the 1D case. Using similar assumptions as in the 1D case, 
 \begin{align}
   \ddot{R} &= \left[ \frac{\rho_{B0} c^2}{\rho_L} \left(\frac{R_{B0}^2}{R^2} -1 \right)  + \frac{\dot{R}^2}{2}\left(1- \frac{R^2}{R_{\infty}^2}  \right) -\dot{R}^2 \ln\frac{R_{\infty}^2}{R}\right] \frac{1}{R ln\left( \frac{ R_{\infty}^2}{R}  \right)} \\
   -\omega^2 r e^{i\omega t} &= \left( -\frac{\rho_{B0}c^2}{\rho_L}\frac{2r'}{R_0}e^{i\omega t} \right) \frac{1}{R_0 \ln V_R} \qquad \textrm{(using $V_R = \frac{R_{\infty 0}}{R_0}$)} \\ 
   \omega^2 &= \frac{2\rho_{B0} c^2}{\rho_L R_0^2 \ln V_R}
 \end{align}
 The natural frequency of the circular RP system is 
 \begin{equation}
   f_N = \frac{\omega}{2\pi} = \frac{1}{2\pi}\sqrt{ \frac{2\rho_{B0}c^2}{\rho_L R_0^2 ln \left(\frac{R_0}{R_{\infty 0}}\right)}}
   \label{2d_frequency}
 \end{equation}
  Figure \ref{fig:2drp} shows the pressure and radius response of the bubble for a 
 density ratio of  10:1. The pressure response is shown in the frequency domain 
 in Fig. \ref{fig:2drpf}. 
 As seen in Fig. \ref{fig:2drp}, the radius seems to be more stabilized and 
 continues for larger time, whereas the pressure in the interior rises and 
 deviates from the theoretical result. Possibly, this is the effect of inaccuracy 
 in the 
 surface tension model applied. At higher density ratios the interface 
 undergoes much severe distortions and leads to mixing of the two fluids (not presented here). 
 However, the initial time response is accurately captured. More accurate 
 surface tension implementations, for example, that proposed in \cite{adami2010new} or \cite{rezavand2018isph}
 for larger density ratio interfaces may be implemented to improve the artificial
 mixing of the phases at the interface. In Fig. \ref{circle_time}, the location 
 of the interface at different times can be seen. At $t=130$ we see that the 
 particles begin get disturbed from their initial lattice. We note that these distortions
 can be reduced by remeshing \cite{chaniotis2002remeshed} or redistributing \cite{xu2009} the particles.

\section{Summary}
Several two phase flow problems of practical importance involve compressibility 
effects 
in at least one of the phases. We coupled compressible and the incompressible SPH 
methods such that the compressible phase provided the Dirichlet boundary
condition for pressure to the incompressible phase. To motivate application to 
gas bubbles moving through strong pressure gradients, we derived the RP 
equation for a rectangular and a circular bubble in two dimensions and compared 
the CI--SPH simulations with these results. A basic surface tension model was used. For
Dirichlet BC at the free surface, a penalty term was used in both the PPE and the pressure gradient 
approximation. 

Artificial mixing of the two phases at the interface was observed. This could be
due to the inaccuracies in the surface tension model. Improved surface tension models could rectify this problem. 
Notwithstanding, simulations for up to a density ratio of 1000:1  
compared well with the theory. 

\section*{Acknowledgements}
The authors gratefully acknowledge the support of the Cluster of Excellence Engineering of
Advanced Materials, ZISC, FPS and the Collaborative Research Center SFB814 funded by the German Science Foundation (DFG),
and the Indo-German Partnership in Higher Education Program (IGP) 2016, funded by 
DAAD-UGC. 


\end{document}